\documentclass{ipi}

\usepackage{amsmath}
\usepackage{paralist}
\usepackage{graphics}
\usepackage{epsfig}
\usepackage{graphicx}
\usepackage{epstopdf}
\usepackage{mathtools}
\usepackage[plain,lined,figure]{algorithm2e}
\usepackage[colorlinks=true]{hyperref}
\hypersetup{urlcolor=blue, citecolor=red}

\def\currentvolume{X}
\def\currentissue{X}
\def\currentyear{20xx}

\def\ppages{X--XX}
\def\DOI{10.3934/ipi.xx.xx.xx}

\newcommand{\dd}[2]{\frac{\mathrm{d}{#1}}{\mathrm{d}{#2}}}

\newcommand{\pd}[2]{\frac{\partial{#1}}{\partial{#2}}}
\newcommand{\pl}[2]{\partial{#1}/\partial{#2}}

\newcommand{\abs}[1]{\lvert#1\rvert}
\newcommand{\norm}[1]{\lVert#1\rVert}

\DeclareMathOperator{\artanh}{artanh}

\title[Poincar{\'e} inverse problem]
{Poincar{\'e} inverse problem and torus construction in phase space}
\author[Teemu Laakso and Mikko Kaasalainen]{}


\email{teemu.laakso@tut.fi}
\email{mikko.kaasalainen@tut.fi}

\begin{document}
\maketitle

\centerline{\scshape  Teemu Laakso and Mikko Kaasalainen}
\medskip
{\footnotesize
\centerline{Department of Mathematics, Tampere University of Technology}
\centerline{PO Box 553}
\centerline{33101 Tampere, Finland}}

\bigskip

\centerline{(Communicated by the associate editor name)}

\begin{abstract}
The phase space of an integrable Hamiltonian system is foliated by invariant tori. For an arbitrary Hamiltonian $H$ such a foliation may not exist, but we can artificially construct one through a parameterised family of surfaces, with the intention of finding, in some sense, the closest integrable approximation to $H$. This is the Poincar{\'e} inverse problem (PIP). In this paper, we review the available methods of solving the PIP and present a new iterative approach which works well for the often problematic thin orbits.   
\end{abstract}

\section{Introduction}
Integrability of a dynamical system, witnessed by a foliation of invariant toroidal manifolds in the phase space, is a highly useful property. For a Hamiltonian $H_0$ which is autonomous and integrable, we have a symplectic map (often explicit through the Hamilton-Jacobi approach)
\begin{equation}
\label{e:torusmap}
(\theta,J)\leftrightarrow(q,p),
\end{equation}
between the action-angle variables $(\theta,J)$ and the Cartesian phase-space coordinates $(q,p)$. Not only does this eliminate the need for any numerical integration of the equations of motion, but also allows one to study the canon of near-integrable Hamiltonians $H$ in a form of power series;
\begin{equation}
\label{e:perturbed_H}
H(\theta,J)=H_0(J)+\varepsilon H_1(\theta,J)+\varepsilon^2 H_2(\theta,J)+\ldots,
\end{equation}
where the strength of the perturbation $\varepsilon$ is small. To study Eq.\ \eqref{e:perturbed_H} and the corresponding equations of motion, according to Poincar{\'e} \cite{Poi1892}, is ``the general problem of dynamics''. We consider it as the direct problem, and ask the inverse: ``given a near-integrable Hamiltonian in the form $H(q,p)$, what is the best way of defining the maps \eqref{e:torusmap} and $J\mapsto H_0(J)$?'', i.e., we are looking for the closest integrable approximation to $H$. Indeed, there are many cases where a system shows near-integrable behaviour numerically, but there is no prior information on $H_0$, underlining the relevance of the question above.

The Poincar{\'e} inverse problem (PIP) is obviously an ill-posed problem (in the sense of Hadamard); particularly, in regions of phase space, where invariant toroidal manifolds are absent, the closeness of $H$ and $H_0$ is an insufficient requirement for a unique solution. Simply minimising $\norm{H-H_0}$ everywhere can lead to unacceptable results \cite{Kaa1994}. Generally, in order to guarantee that the solution is a reasonable approximation to the actual motion, additional constraints (regularisation) are necessary. On the other hand, since $H$ is a known function, the data are error-free sampling points, and their number is only limited by the available computational resources. Therefore, ``model noise'' is the fundamental source of error, which is typical for many inverse problems. 

The PIP can also be a subproblem of a larger inverse one. For example, in dynamical or phase-space tomography \cite{Kaa2008} we want to solve for a potential field $\Phi$ and phase-space distribution functions best corresponding to observed positions and velocities of matter. In this process, we repeatedly construct an $H_0$ corresponding to a trial $\Phi$ in order to describe the phase-space distribution of observed matter along the tori of $H_0$.

Our approach towards the solution of the PIP is based on numerical construction of phase-space tori which should coincide with the existing Kolmogorov-Arnold-Moser (KAM) tori of $H$ (if any); i.e., on these, $\norm{H-H_0}=0$ within the available accuracy. This ensures that the tori of $H=\mathcal{H}(\mathcal{J})$, where $\mathcal{H}$ is any specific integrable Hamiltonian and $\mathcal{J}$ are its actions, are faithfully reconstructed. Also, for a perturbed $H=\mathcal{H}+\varepsilon\mathcal{H}_1+\varepsilon^2\mathcal{H}_2+\ldots$ of the form \eqref{e:perturbed_H}, we have the natural limit
\begin{equation*}
\lim_{\varepsilon\rightarrow 0} H_0(J;\varepsilon)=\lim_{\varepsilon\rightarrow 0} H(q,p;\varepsilon)=\mathcal{H}(\mathcal{J}).
\end{equation*}
Note that it is not necessary to have $H_0=\mathcal{H}$, when $\varepsilon>0$. Even here there is some ambiguity; a torus can be analytically defined as a KAM one only by analysing it via the perturbation \eqref{e:perturbed_H}; whether we can directly deem an orbit of $H$ to correspond to a KAM torus is usually up to numerical resolution. As implied by the above principles, our approach is geometric rather than dynamical (which was the line of thought preferred by Poincar{\'e} as well).

We construct the tori one by one, and shall refer to this process as torus construction, torus modelling, or torus embedding. The full solution, defining the maps \eqref{e:torusmap} and $J\mapsto H_0(J)$ everywhere, requires an interpolation scheme between the constructed tori. What interpolations produce a foliated global set of tori that really correspond to an $H_0$ is an open question. We shall not attempt to answer it in this paper; instead, we concentrate on the fundamental geometric problem of defining the local tori. This part of the PIP reduces to the problem of surface reconstruction, if only the KAM tori are modelled. However, this is generally not sufficient; when $H$ is far from integrable, its set of KAM tori may be too sparse to form an interpolation grid. One may even have a case where $H$ has no KAM tori at all \cite{Kaa1995}. Hence, the ability of embedding approximating tori in the absence of KAM ones is also of practical importance, and it shall be a priority in our approach.

This paper is organised as follows. In Sect.\ \ref{s:overview} we review and classify some of the available methods of torus construction, and in Sect.\ \ref{s:model} we introduce a new one. In Sect.\ \ref{s:examples:1d} and \ref{s:examples:2d} we present numerical examples and discuss some practical choices during the implementation. In Sect.\ \ref{s:discussion} we sum up and discuss the performance of the method.

\section{Overview of methods}
\label{s:overview}
Two important questions that characterise any approach to torus construction are: a) how to define and parametrize a model torus, i.e.,\ the map \eqref{e:torusmap}, and b) how to fit it in the phase space. In order to guarantee periodicity in the angle coordinates $\theta$, a typical, and natural, torus model involves Fourier series approximation. The approximated quantities can be simply the phase-space coordinates in the form $q(\theta)$ and $p(\theta)$ (\emph{direct model}), or alternatively, the map \eqref{e:torusmap} can be carried through using the action-angle variables $(\vartheta,\mathcal{J})$ of some known integrable Hamiltonian $\mathcal{H}$ as an intermediate step. In this case, a Fourier series is used to define a canonical transformation $(\theta,J)\leftrightarrow(\vartheta,\mathcal{J})$ by a generating function (\emph{GF model}). 

Among the methods of fitting the model torus to the Hamiltonian flow of $H$, a characterising feature is whether phase-space quantities are sampled along numerically integrated orbits (\emph{trajectory method}), or are they iteratively adjusted on a grid of angles (\emph{iterative method}). Worth noting are also the variables ($J$, $(q,p)$, or the frequencies $\omega$) which \emph{label}, i.e., define along with the Fourier coefficients, the torus to be modelled.

A diverse selection of methods have been developed, each with their own strengths and weaknesses. Valluri \& Merritt \cite{VM1999} give a more thorough review of the subject. 

For non-resonant regular orbits, direct trajectory methods, where the fundamental frequencies are determined from the Fourier spectra (e.g., \cite{BS1982,ME1987,PL1996}), are a robust choice. Having been refined by many authors, they show good accuracy which, however, decays near resonances, where it takes longer for a sample orbit to fill the torus. In addition, since chaotic orbits do not lie on tori at all, they are a questionable target for all trajectory methods. The hybrid GF-method of Warnock \cite{War1991} is efficient, but also affected by the limitations above.

For solving the PIP, it is important that tori can be embedded everywhere in the phase space. In this respect, the iterative method of McGill \& Binney \cite{MB1990} with its further refinements \cite{BK1993,KB1994a,LK2013} stands out, because it is unaffected by resonances, and can construct approximations to invariant tori also in phase-space regions where KAM tori do not exist \cite{Kaa1994,Kaa1995}. As a GF-model it has the advantage that the constructed tori are inherently canonical. On the other hand, it relies on the existence of an integrable $\mathcal{H}$ whose tori resemble those of $H$. The wide class of St{\"a}ckel Hamiltonians is available for this purpose, but at the cost of increased analytical and computational effort \cite{LK2013}. 

The use of simpler, but more restricted $\mathcal{H}$ emphasises a general limitation of GF models; their inability to produce orbits which are arbitrarily thin; i.e., orbits for which $J_\rho\rightarrow 0$, where $J_\rho$ is the generalised transversal action, describing the thickness of the orbit in the configuration space \cite{KB1994a,Laa2011}. This happens whenever orbits of $H$ and $\mathcal{H}$ with $J_\rho=0$ do not align in configuration space, which is usually the case. In two-dimensional systems in configuration space, this can sometimes be circumvented by using (rather cumbersome) point transformations \cite{KB1994a}. In three dimensions, this becomes much more difficult, and in any case there are tori in systems of all dimensions that cannot be modelled via such transformations. This is an important point when considering the general applicability of a method; indeed, we argue that only the type of method presented in this paper can be called generally applicable. 

The main problem with point transformations is that they are only based on the shape of the orbit in configuration space. Thus they separate, e.g., librating and circulation motion explicitly into two topologically different categories though in phase space they share the same topology. Furthermore, the point transformations are used to describe a coordinate system around the orbit $J_\rho=0$, in which one hopes to be able to describe the orbits $J_\rho\approx 0$ in a consistent way. In general, there is no guarantee that such a system exists; for example, some orbits in a rotating potential \cite{Kaa1995} are obviously not describable in this manner. In some cases, the image of the orbit in such a system becomes extremely complicated. Also, in two dimensions, the orbits at $J_\rho=0$ are closed and thus easily integrated, whereas in three dimensions orbits for which $J_\rho=0$ describe a surface in configuration space. Constructing this surface to define the desired coordinate system is more difficult than determining a curve.

In this paper, acknowledging the limitations related to trajectory and GF-methods above, we propose a new approach which is direct and iterative. By doing this, we abandon the automatic canonicity of the model, which must be compensated by introducing additional optimisation constraints. A somewhat similar approach was taken by Ratcliff et al. \cite{RCS1984}, but the present algorithm is less restricted and more robust, since the model is fitted in the least-squares sense, and physically relevant optimisation functions are used. What is more, we introduce a principle for creating tori by letting model tori to adjust themselves by ``floating''; i.e., their labels themselves are let to wander locally to some extent, which gives additional flexibility. We illustrate the new method by applying it to selected one- and two-dimensional systems commonly used in galaxy modelling.

\section{The torus model}
\label{s:model}
We represent the model torus as a Fourier series of the Cartesian phase-space coordinates $q\in\mathbb{R}^n$ and velocities $p\in\mathbb{R}^n$;
\begin{equation}
\label{e:torusmap:pq}
p(\theta)=\sum_{k\in A}\alpha_ke^{i(k\cdot\theta)},\qquad q(\theta)=\sum_{l\in B}\beta_le^{i(l\cdot\theta)},
\end{equation}
where $A,B\subset\mathbb{Z}^n$ are sets of multi-indices, $\alpha_k,\beta_l\in\mathbb{C}^n$ are Fourier coefficients, and $\theta\in[0,2\pi)^n$ are canonical angles, conjugate to some actions $J_i\ge 0$, $i=1,\ldots,n$ which label the model torus. The real-valuedness of $p$ and $q$ dictates that  $\alpha_{-k}=\overline{\alpha}_k$ and $\beta_{-l}=\overline{\beta}_l$. The strict consistency of the Cartesian momenta, $p\equiv\dot{q}$, would fix half of the coefficients: $A=B$, $\alpha_k=i(k\cdot\omega)\beta_k$. However, at this point, we wish to retain the maximal flexibility of the model (as a general $\theta$-parameterised surface in the $pq$-space), and do not enforce this identically.

In order to find the optimal $\alpha_k$ and $\beta_l$ for a given Hamiltonian $H:\mathbb{R}^n\times\mathbb{R}^n\to\mathbb{R}$, $(q,p)\mapsto H(q,p)$, we build an objective function $R$ which is a weighted sum
\begin{equation}
\label{e:objective}
R(\theta)=\sum_i\lambda_i\norm{\mathcal{E}_i(\theta)}^2
\end{equation}
of error functions $\mathcal{E}_i$ that vanish on a KAM torus of $H$. The norm is Euclidean, and $\lambda_i>0$. We will minimise $R$ on a grid of angles $\theta_{(m)}$, $m=1,\ldots,M$ with respect to the Fourier coefficients $\alpha_k,\beta_l$. Obviously, success in torus construction requires that the error functions $\mathcal{E}_i$ are chosen wisely. This is our goal in the following.

Hamilton's equations of motion under $H$ are
\begin{equation}
\label{e:hem:pq}
\dot{p}=-\partial_qH,\quad\dot{q}=\partial_pH,
\end{equation}
and supposing that $(\theta,J)$ are action-angle variables of $H$, we also have
\begin{equation}
\label{e:hem:Jtheta}
\dot{J}=0,\quad\dot{\theta}=\partial_JH=\omega.
\end{equation}
Applying the chain rule to the time derivatives of the model \eqref{e:torusmap:pq}, inserting \eqref{e:hem:Jtheta} and equating to \eqref{e:hem:pq} yields
\begin{equation}
\label{e:hem:combined}
\pd{p}{\theta}\omega=-\pd{H}{q},\quad\pd{q}{\theta}\omega=\pd{H}{p}
\end{equation}
which identifies the vector field produced by the model with the Hamiltonian vector field of $H$. Assuming that $H$ is differentiable, the partial derivatives in Eq.\ \eqref{e:hem:combined} can be evaluated for all $\theta$. We define two error functions,
\begin{equation}
\label{e:errorf:12}
\mathcal{E}_1(\theta)\coloneqq\pd{p}{\theta}\omega+\pd{H}{q},\quad\mathcal{E}_2(\theta)\coloneqq\pd{q}{\theta}\omega-\pd{H}{p},
\end{equation}
which penalise for any deviation from the Hamiltonian flow of $H$. Obviously, in order to use Eq.\ \eqref{e:errorf:12} we need the values of the frequencies $\omega$. A simple solution would be to use $\omega$ for labelling the model torus, i.e., to use $\omega$ as constants. However, sets of valid frequencies can be difficult to sample, and they may overlap for different orbit families, which makes $J$ the preferred label.

Actually, we can avoid labelling the model torus (beforehand) altogether, if we dictate that $\omega$ is a least-squares estimator, obtained by fitting the $2n$ equations \eqref{e:hem:combined} at all of the grid points. More specifically, we have an overdetermined linear system of the form $A\omega=b$, where $A$ is an $(2nM\times n)$-matrix obtained by concatenating the $(n\times n)$-matrices $\pl{p}{\theta_{(m)}}$ and $\pl{q}{\theta_{(m)}}$ for all $m=1,\ldots,M$. The solution is given by the normal equations;
\begin{equation}
\label{e:omega:model}
\omega=(A^TA)^{-1}A^Tb.
\end{equation}
This complicates the dependence of $\mathcal{E}_1$ and $\mathcal{E}_2$ on the Fourier coefficients somewhat, but proved to be a numerically robust solution in our experiments.

Convergence in minimisation towards a KAM torus of $H$ is possible with just the two error functions \eqref{e:errorf:12}.  This principle was employed by Ratcliff et al.\ \cite{RCS1984}, with the aim of having the two vanish at a given number of phase-space points. This, however, causes uncontrollable fluctuations on the rest of the constructed torus when it does not approximate a KAM one of $H$. Our reconstruction is based on the geometric principle of least-squares fitting a number of error functions $\mathcal{E}_i$ that are equivalent on KAM tori (subsets of them are necessary and sufficient), but different elsewhere. In other words, we construct the desired manifolds by regularisation and global smoothing.

From  Hamilton's equations of motion \eqref{e:hem:Jtheta} we can easily derive
\begin{equation}
\label{e:errorf:3}
\mathcal{E}_3(\theta)\coloneqq\dd{J}{t}=\pd{H(q(\theta),p(\theta))}{\theta}=\pd{H}{q}\pd{q}{\theta}+\pd{H}{p}\pd{p}{\theta}
\end{equation}
which gives additional support to, and falls into the same category as, $\mathcal{E}_1$ and $\mathcal{E}_2$. In numerical exercises, where Eq.\ \eqref{e:omega:model} was used, we found that the inclusion of $\mathcal{E}_3$ improved our results noticeably.

Further regularisation is probably necessary, at least when constructing tori in non-regular regions of phase space of $H$. A useful error function is obtained by dictating that the value of the Hamiltonian $H$ should be constant on the minimisation grid;
\begin{equation}
\label{e:errorf:4}
\mathcal{E}_4(\theta)\coloneqq H(\theta)-\bar{H},
\end{equation}
where
$\bar{H}$ is the arithmetic mean of $H$ over the grid. This also yields the local solution of PIP: we define $H_0:=\bar{H}$ on the constructed torus. The global interpolation scheme over the tori, must, of course, fulfill $\omega(J)=\partial H_0/\partial J$ everywhere.

Finally, we introduce an error function which ensures that the model converges towards a specific torus with given values of $J$. For this purpose, we need to derive expressions for the model actions. By definition,
\begin{equation*}
J_h=\frac{1}{2\pi}\oint_{\gamma_h}p\cdot\mathrm{d}q,\quad h=1,\ldots,n,
\end{equation*}
where $\gamma_h$ is a closed path that cannot be continuously deformed into a point. For a model of the form $q(\theta)$ we have
\begin{equation*}
\mathrm{d}q=\sum_{j=1}^n\pd{q}{\theta_j}\mathrm{d}\theta_j,
\end{equation*}
and if we select the path $\gamma_h$ in such a way that $\mathrm{d}\theta_j=0$ for $j\neq h$, we can write 
\begin{equation*}
J_h=\frac{1}{2\pi}\sum_{j=1}^n\int_0^{2\pi}p_j\pd{q_j}{\theta_h}\mathrm{d}\theta_h.
\end{equation*}
By inserting the model equations \eqref{e:torusmap:pq} and integrating, we obtain
\begin{equation}
\label{e:J:model}
J_h=\sum_{j=1}^n\sum_{(k,l)\in D_h}ik_h\alpha_{j,k}\beta_{j,l}e^{i\left[(k+l)\cdot\theta\right]},
\end{equation}
where $D_h=\{(k,l)\in A\times B:l_h=-k_h\}$, and $\alpha_{j,k}$ and $\beta_{j,l}$ are the $j$:th components of the corresponding Fourier coefficients. Note that $J_h$ does not depend on $\theta_h$. On a KAM torus, the dependence on $\theta_j$, $j\neq h$ should vanish as well. In order to enforce this, and to punish for any deviation from a given set of actions $J=\bar{J}$, we introduce an error function
\begin{equation}
\label{e:errorf:5}
\mathcal{E}_5(\theta)\coloneqq J(\theta)-\bar{J}.
\end{equation}
As our notation suggests, $\bar{J}$ could also be computed as an arithmetic mean over the grid, supplementing the label-free version of the torus construction algorithm.

\section{Numerical example: isochrone potential}
\label{s:examples:1d}
Let us first check how the method reconstructs simple integrable Hamiltonians. The harmonic oscillator is trivial, since it is given directly in the Fourier form of Eq.\ \eqref{e:torusmap:pq} with only first-order terms. As a more interesting sanity check for the presented torus model and minimisation criteria, we apply them to the isochrone Hamiltonian \cite{Hen1959} with a single degree of freedom;
\begin{equation*}
H(q,p)=\frac{1}{2}p^2-\frac{c_1}{c_2+\sqrt{c_2^2+q^2}},
\end{equation*}
where $q,p\in\mathbb{R}$, and the parameters have values $c_1=1$ and $c_2=0.15$. By fixing the origin of $\theta$ to $q=0$, and acknowledging the symmetries of the system, we express the torus model \eqref{e:torusmap:pq} in trigonometric form;
\begin{equation*}
p(\theta)=\sum_{k\in X}a_k\cos(k\theta),\qquad q(\theta)=\sum_{l\in Y}d_l\sin(l\theta),
\end{equation*}
where $X=Y=\{1,3,5,7,\ldots,N-1\}$. Since the system is one-dimensional and integrable, we can, in this case, safely label the model tori by their frequencies $\omega$, which simplifies the error functions \eqref{e:errorf:12}. We have
\begin{align*}
\mathcal{E}_1(\theta)&=\omega p'(\theta)+\frac{c_1q(\theta)}{\sqrt{c_2^2+q^2(\theta)}(c_2+\sqrt{c_2^2+q^2(\theta)})},\\
\mathcal{E}_2(\theta)&=\omega q'(\theta)-p(\theta),
\end{align*}
where
\begin{equation*}
p'(\theta)=-\sum_{k\in X}ka_k\sin(k\theta),\qquad q'(\theta)=\sum_{l\in Y}ld_l\cos(l\theta).
\end{equation*}
The rest of the error functions $\mathcal{E}_i$, $i=3,4,5$ will not be used in this example. We minimise the objective function \eqref{e:objective} (with $\lambda_1=\lambda_2=1$ ) using the Levenberg-Marquardt algorithm (LMA). This requires the calculation of the partial derivatives $\pl{\mathcal{E}_i}{a_k}$ and $\pl{\mathcal{E}_i}{d_l}$, $i=1,2$, which is straightforward using the equations above.

\begin{figure}[htb]
\includegraphics[width=\columnwidth]{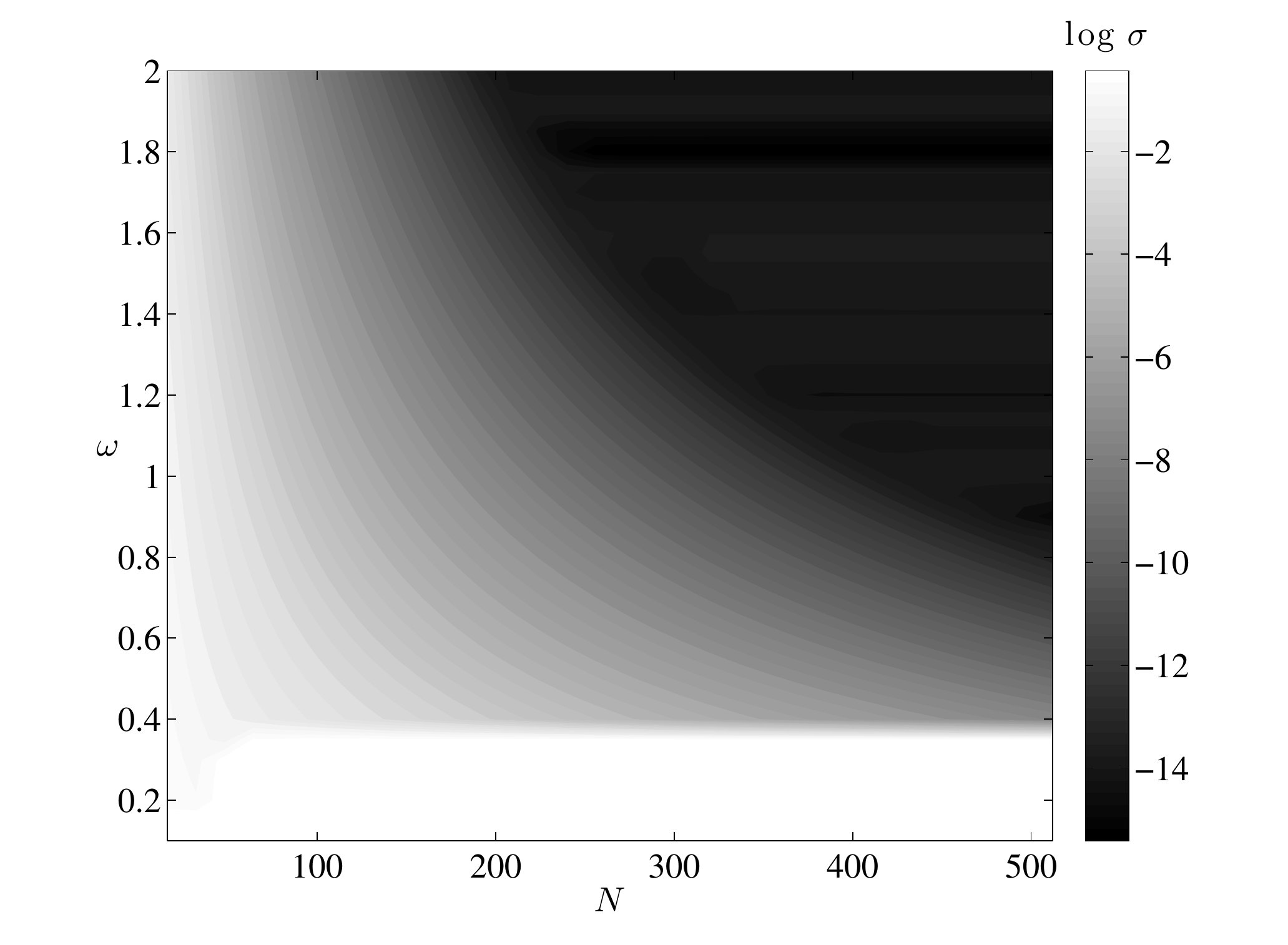}
\caption{Standard deviation $\sigma$ of the one-dimensional isochrone Hamiltonian function values on the $\theta$-grid after running the torus embedding LMA.}
\label{f:ic:N:omega}
\end{figure}

With the purpose of studying the convergence radius and accuracy of the method, we ran the LMA several times, while varying the number of coefficients $N=16,32,48,\ldots,512$ and the frequency $\omega=0.1,0.15,0.2,\ldots,2.0$. The angle grid contained $M=1024$ equally spaced values (matching the Nyquist rate for $N=512$), but due to the symmetry of the model, only half of them were actually needed and used. As an initial guess for the model torus we set all the coefficients to zero except $a_1=d_1=1$, which corresponds to a unit circle in the $qp$-plane (i.e., a harmonic oscillator). For each pair $(N,\omega)$, after the LMA was well converged, we evaluated the standard deviation $\sigma$ of the Hamiltonian $H(q(\theta),p(\theta))$ over the grid points. Figure \ref{f:ic:N:omega} shows a contour plot of the results.

For all $\omega\ge 0.4$ the LMA converged, and $H$ attained a constant value throughout the grid, within the accuracy seemingly bestowed by $N$ and also $\omega$. In the upper right portion of the figure, the values of $\sigma$ are saturated due to the finite computational precision (64-bit floating point). For the isochrone, we know the relation $H=-(2c_1\omega)^{2/3}/2$, and hence, we could verify that Fig.\ \ref{f:ic:N:omega} also closely represents the deviation from the analytical values $H(\omega)$. For $\omega<0.4$ there was no proper convergence. However, this could be remedied by compensating the increased amplitude of the motion in the initial guess ($\omega\to 0\implies J\to\infty$). For example, with an initial guess $q(\theta)=2\sin\theta$, $p(\theta)=2\cos\theta$ the LMA converged nicely for $\omega=0.2,\ldots,0.4$. We expect that a decent initial guess is increasingly important for more complex systems.

\section{Numerical example: logarithmic potential and the PPS}
\label{s:examples:2d}
We raise the degrees of freedom, and formulate the torus modelling algorithm for planar systems of the form
\begin{equation*}
H(q,p)=\frac{1}{2}p\cdot p+\Phi(q),
\end{equation*}
where $q,p\in\mathbb{R}^2$ and $\Phi:\mathbb{R}^2\to\mathbb{R}$ is a potential function. Depending on $\Phi$, there may be structural features in phase space, such as differentiation into orbit families, and stochastic regions surrounding resonant islands (for near-integrable potentials). In order to obtain a basic understanding how the torus algorithm copes with these features, we apply it to two selected potentials. First, we have the near-integrable logarithmic potential \cite{Ric1980,BT1987}
\begin{equation}
\label{e:potential:logarithmic}
\Phi(q)=\frac{1}{2}\ln\left(q_1^2+\frac{q_2^2}{c_1^2}+c_2^2\right)
\end{equation}
where $c_1=0.9$ and $c_2=1$. The phase space of the logarithmic system is dominated by KAM tori, and with the selected parameter values, most of the orbits are either boxes (i.e.\ butterflies) or loops \cite{MS1989}.

For comparison, as an integrable example, we use a St{\"a}ckel potential in elliptic coordinates $u$;
\begin{equation}
\label{e:potential:stackel}
\Phi(u)=-\frac{f(u_1)-f(u_2)}{u_1-u_2},
\end{equation}
where $f$ is a smooth function. The coordinate transformation $q\mapsto u$ is given by 
\begin{equation*}
u_1+u_2=-c_1-c_2+q_1^2+q_2^2,\qquad u_1u_2=c_1c_2-c_2q_1^2-c_1q_2^2,
\end{equation*}
where $c_1<c_2$ are parameters. By selecting $u_2\le u_1$, we have $-c_2\le u_2\le -c_1\le u_1$. With the choice
\begin{align*}
f(u_1)&=-2\pi(u_1+c_1)c_2c_3\sqrt{\frac{-c_1}{u_1+c_1}}\arctan\sqrt{\frac{u_1+c_1}{-c_1}},\\
f(u_2)&=-2\pi(u_2+c_1)c_2c_3\sqrt{\frac{c_1}{u_2+c_1}}\artanh\sqrt{\frac{u_2+c_1}{c_1}},
\end{align*}
the potential \eqref{e:potential:stackel} becomes the perfect prolate spheroid (PPS), a special case of the perfect ellipsoid \cite{Zee1985}. We set the parameter values to $c_1=-1$, $c_2=-0.25$, and $c_3=1$. In the $q_1q_2$-plane, the PPS hosts the same major orbit families, the boxes and loops, as the logarithmic potential \eqref{e:potential:logarithmic}. In fact, the phase-space structures of the two systems are very similar, and in a side-by-side comparison of the corresponding torus models, integrability of the PPS supports the analysis on the logarithmic side as well.

\begin{figure}[htb]
\includegraphics[width=\columnwidth]{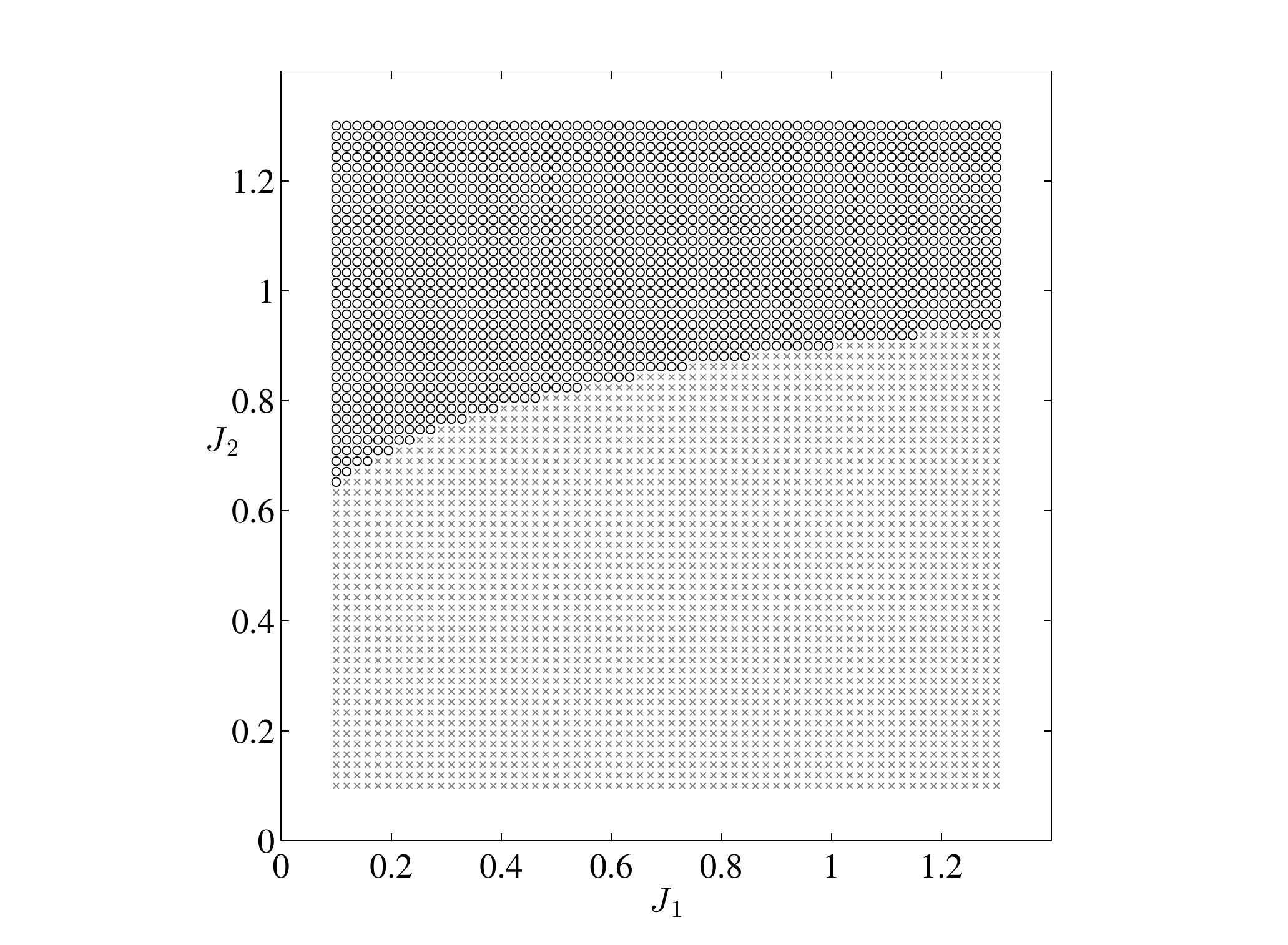}
\caption{The loops (black circles) and boxes (gray crosses) in the action space of the PPS.}
\label{f:pps:families:J}
\end{figure}

\begin{figure}[htb]
\includegraphics[width=\columnwidth]{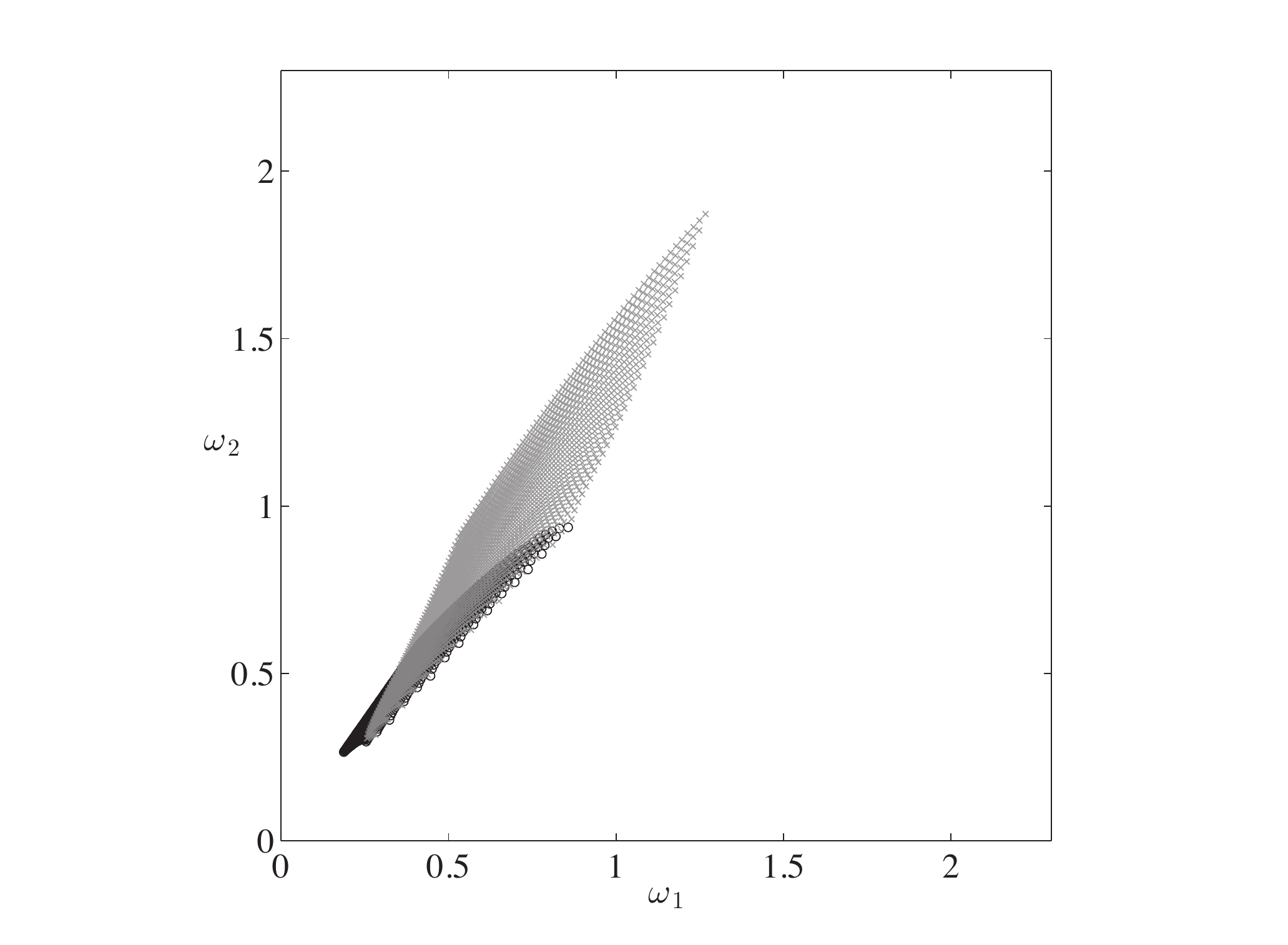}
\caption{The orbits of Fig.\ \ref{f:pps:families:J} mapped into the PPS frequency space.}
\label{f:pps:families:omega}
\end{figure}

Since we expect to model only KAM tori, our main concern is how the orbit family, or the shape of an orbit within the family, affects the outcome of the torus modelling. Compared to the one-dimension case, a big difference is that we now use the actions $J$ to label the tori. There are two reasons for not using $\omega$. First, for the PPS (and thus possibly for the logarithmic system as well), the values of $\omega$ overlap for boxes and loops (Fig.\ \ref{f:pps:families:J} and \ref{f:pps:families:omega}). Second, in some of our experiments while using $\omega$ as a label, the LMA converged to a trivial $1$-torus ($J_1=0$ or $J_2=0$). 

A general trigonometric version of the torus model \eqref{e:torusmap:pq} is now
\begin{equation}
\label{e:torusmap:pq:2}
\begin{aligned}
p(\theta)&=\sum_{k\in X}a_k\cos(k\cdot\theta)+b_k\sin(k\cdot\theta),\\
q(\theta)&=\sum_{l\in Y}c_l\cos(l\cdot\theta)+d_l\sin(l\cdot\theta),
\end{aligned}
\end{equation}
where $q,p\in\mathbb{R}^2$, $a_k,b_k,c_l,d_l\in\mathbb{R}^2$, $\theta\in[0,2\pi)^2$, and $X,Y\subset\mathbb{Z}^2$ are sets which do not contain opposite indices. Due to symmetries and other characteristics of a given orbit (family), some terms in Eq.\ \eqref{e:torusmap:pq:2} are always zero. Significant savings in computation time of the LMA are obtained, if the corresponding indices are identified and omitted from $X$ and $Y$. For the integrable PPS there is a shortcut for doing this numerically. Namely, since we know the transformation $(\theta,J)\mapsto(q,p)$, we can compute sample points $q(\theta)$ and $p(\theta)$ on an even $\theta$-grid, and perform a discrete Fourier transformation (DFT) which yields a representation exactly of the form of our model. The computation is relatively fast (with optimised FFT-routines), and the DFT orbit can also be used to verify the result of the LMA. For the PPS box orbits, we found the nonzero coefficients to be
\begin{equation}
\label{e:coefficients:box}
\begin{aligned}
&a_{1,k},\text{ if }k=(\text{odd, even}),\\
&a_{2,k},\text{ if }k=(\text{even, odd}),
\end{aligned}
\qquad
\begin{aligned}
&d_{1,l},\text{ if }l=(\text{odd, even}),\\
&d_{2,l},\text{ if }l=(\text{even, odd}),
\end{aligned}
\end{equation}
and for the loops; 
\begin{equation}
\label{e:coefficients:loop}
\begin{aligned}
&b_{1,k},\text{ if }k=(\text{even, odd}),\\
&a_{2,k},\text{ if }k=(\text{even, odd}),
\end{aligned}
\qquad
\begin{aligned}
&c_{1,l},\text{ if }l=(\text{even, odd}),\\
&d_{2,l},\text{ if }l=(\text{even, odd}).
\end{aligned}
\end{equation}
We define two separate initial guesses for the model \eqref{e:torusmap:pq:2}; one for box orbits, and the other for loops, containing only coefficients of the form \eqref{e:coefficients:box} and \eqref{e:coefficients:loop}, respectively. Because of the common orbit families, these initial guesses work for both the PPS and the logarithmic potential.

Let $N\in\mathbb{N}$ limit the range of the indices such that $\abs{k_i}\le N$, $\abs{l_i}\le N$, $i=1,2$.  By setting $N=16$, we end up with a total of $544$ coefficients per orbit which are initially set to zero except the ones appearing below. As an initial box orbit, we use independent harmonic oscillators in both coordinate directions;
\begin{equation}
\label{e:initial:q:box}
q(\theta)=
\renewcommand{\arraystretch}{1.4}
\begin{pmatrix}
\displaystyle
\sin\theta_1\\
\displaystyle
\sin\theta_2
\end{pmatrix},\qquad
p(\theta)=
\renewcommand{\arraystretch}{1.4}
\begin{pmatrix}
\displaystyle
\cos\theta_1\\
\displaystyle
\cos\theta_2
\end{pmatrix},
\end{equation}
which corresponds to $\omega=(1,1)$. For a loop orbit we must include more nonzero terms, in order to have the initial $J_1>0$ in Eq.\ \eqref{e:J:model}. After some experiments, we found that
\begin{equation}
\label{e:initial:q:loop}
q(\theta)=
\renewcommand{\arraystretch}{2}
\begin{pmatrix}
\displaystyle
\cos\theta_2+\frac{1}{20}\cos(2\theta_1+\theta_2)-\frac{1}{2}\cos(-2\theta_1+\theta_2)\\
\displaystyle
\frac{3}{2}\sin\theta_2+\frac{1}{10}\sin(2\theta_1+\theta_2)-\frac{1}{2}\sin(-2\theta_1+\theta_2)
\end{pmatrix},
\end{equation}
and the $p(\theta)$ that follows from selecting $\omega=(\frac{1}{2},\frac{1}{2})$ are adequate.

Using the least-squares frequencies \eqref{e:omega:model}, the error functions $\mathcal{E}_1$ and $\mathcal{E}_2$ in Eq.\ \eqref{e:errorf:12} for all grid points are given by $f(\theta_{(m)},m=1,\ldots,M)=A\omega-b$. Differentiation with respect to the model coefficients gives
\begin{equation}
\label{e:pd:f:x}
\pd{f}{x}=A(A^TA)^{-1}\pd{A^T}{x}(b-A\omega)+(A(A^TA)^{-1}A^T-I)\left(\pd{b}{x}-\pd{A}{x}\omega\right),
\end{equation}
where $x=a_k,b_k,c_l,d_l$. Since the partial derivatives required for $\pl{\mathcal{E}_3}{x}$ and $\pl{\mathcal{E}_4}{x}$ already appear in Eq.\ \eqref{e:pd:f:x}, there is no trouble in adding $\mathcal{E}_3$ and $\mathcal{E}_4$ to the objective function \eqref{e:objective}. In order to use $\mathcal{E}_5$ we must evaluate the actions \eqref{e:J:model} for the trigonometric version of the model. Taking care of the indexing, we have, for any $n$, 
\begin{equation}
\label{e:J:model:n}
\begin{split}
J_h=-\frac{1}{2}\sum_{j=1}^n\sum_{(k,l)\in Z_h}k_h\Bigl\{(&a_{j,k}d_{j,l}+b_{j,k}c_{j,l})\cos\left[(k+l)\cdot\theta\right]\\
+(&b_{j,k}d_{j,l}-a_{j,k}c_{j,l})\sin\left[(k+l)\cdot\theta\right]\Bigr\},
\end{split}
\end{equation}
where $Z_h=\{(k,l)\in X\times Y\cup X\times-Y:l_h=-k_h\}$.

We set the $\theta$-grid to be a $32\times 32$ square lattice. This fulfils the requirements of the Petersen-Middleton theorem \cite{PM1962} for the chosen number of model coefficients (i.e.,\ there should be no aliasing). However, due to model symmetries with respect to both coordinate axes, we can omit everything but the values $\theta\in[0,\pi)^2$. This leaves us with only $256$ grid points per orbit. Nevertheless, compared to the one-dimensional case, the computational effort to construct a torus is now far greater. Therefore, we do not try to optimise near to machine precision in the LMA. Instead, we set modest goals for the objective function value and concentrate on finding out whether the algorithm converges towards the correct solution or not.

\subsection{Unlabelled torus construction}
In our first set of experiments, we use an objective function \eqref{e:objective} which is unlabelled, consisting of the error functions $\mathcal{E}_i$, $i=1,\ldots,4$ with the weights $\lambda_i$ set to unity. We run the LMA twice for each system (the PPS and logarithmic); first using the fixed initial guess for box orbits \eqref{e:initial:q:box} and then using the one for loops \eqref{e:initial:q:loop}. In all four cases, the LMA converges fast during the first few iteration steps, before reaching a plateau of slow convergence. During the phase of slow convergence, there may still be apparent changes in the model parameters. We interpret this behaviour being caused by the algorithm first finding a valley of KAM tori, and then advancing along its bottom. Since we have not specified a label, any KAM torus will suffice, and we tune the LMA to stop as soon as the bottom of the valley is reached. For comparing LMA results we again use $\sigma$; the standard deviation of the corresponding Hamiltonian function $H$ over the $\theta$-grid. Table \ref{t:unlabelled} lists the final values of the actions, frequencies, and the values of $\sigma$.
\begin{table}
\caption{Initially unlabelled constructed tori for the PPS and the logarithmic potential.}
\centering
\begin{tabular}{llccc}
& & J & $\omega$ & $\log\sigma$ \\
PPS & box & $(0.19,0.34)$ & $(0.97,1.30)$ & $1\times 10^{-5}$ \\
& loop & $(0.14,1.23)$ & $(0.43,0.60)$ & $8\times 10^{-5}$ \\
logarithmic & box & $(0.16,0.22)$ & $(0.78,0.87)$ & $6\times 10^{-7}$ \\
 & loop & $(0.11,0.76)$ & $(0.58,0.67)$ & $2\times 10^{-6}$ \\
\end{tabular}
\label{t:unlabelled}
\end{table}

\begin{figure}[htb]
\includegraphics[width=\columnwidth]{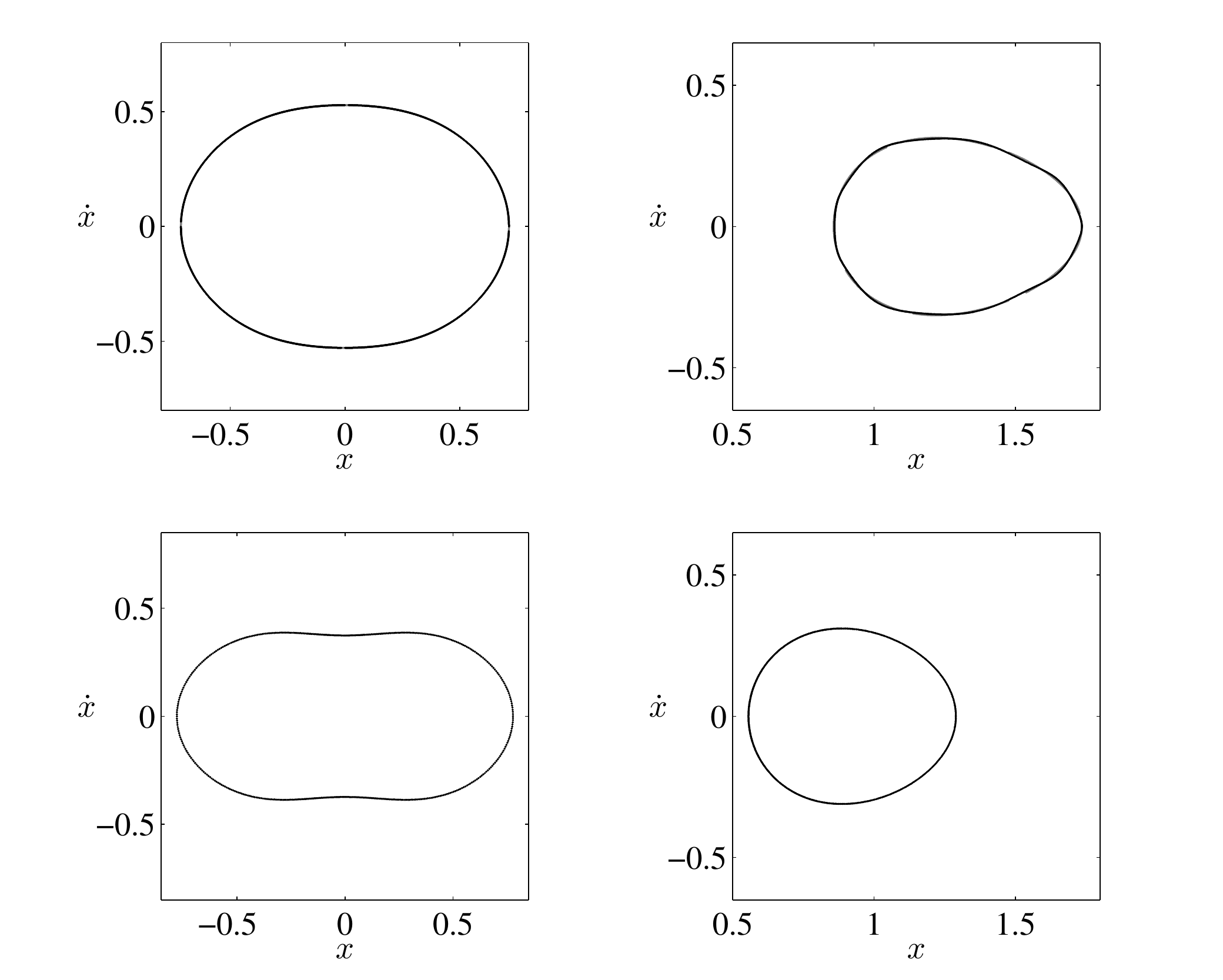}
\caption{Poincar{\'e} sections $y=0$, $\dot{y}>0$ for the unlabelled torus construction, superimposed over numerically integrated reference orbits (gray): the PPS box (upper left) and loop (upper right), and logarithmic box (lower left) and loop (lower right).}
\label{f:unlabelled:poincare}
\end{figure}

It seems that the logarithmic tori are easier to construct than the PPS ones. For both systems, the constructed boxes are more accurate than the corresponding loops. It can be verified that the actions and frequencies in Table \ref{t:unlabelled} all correspond to invariant tori. For the PPS we can do this analytically (cf.\ Fig.\ \ref{f:pps:families:J}), and for both systems we can visually compare constructed orbits to numerically integrated ones, when the orbits share the same initial values. Fig.\ \ref{f:unlabelled:poincare} shows the Poincar{\'e} section $y=0$, $\dot{y}>0$ for the orbits in Table \ref{t:unlabelled}. The common initial point is $\theta=(0,\pi/2)$ on the constructed torus. The numerical integration of Hamilton's equations of motion in Cartesian coordinates is done using the Gragg-Bulirsch-Stoer (GBS) method with high accuracy settings (the standard deviation of $H$ along the integrated orbits is below $10^{-12}$). 
The plotted orbits coincide in all four cases. Hence, the tori are indeed close to invariant.

For the PPS, increasing the number terms by setting $N=24$ (total of $1200$ coefficients), and resizing the $\theta$-grid accordingly, resulted in less than an order of magnitude gain in the final $\sigma$. For our purposes, this was insignificant compared to the increased computation time, and we decided to stick with $N=16$ in the following.

\subsection{Labels from an action grid}
\begin{figure}[htbp]
\includegraphics[width=\columnwidth]{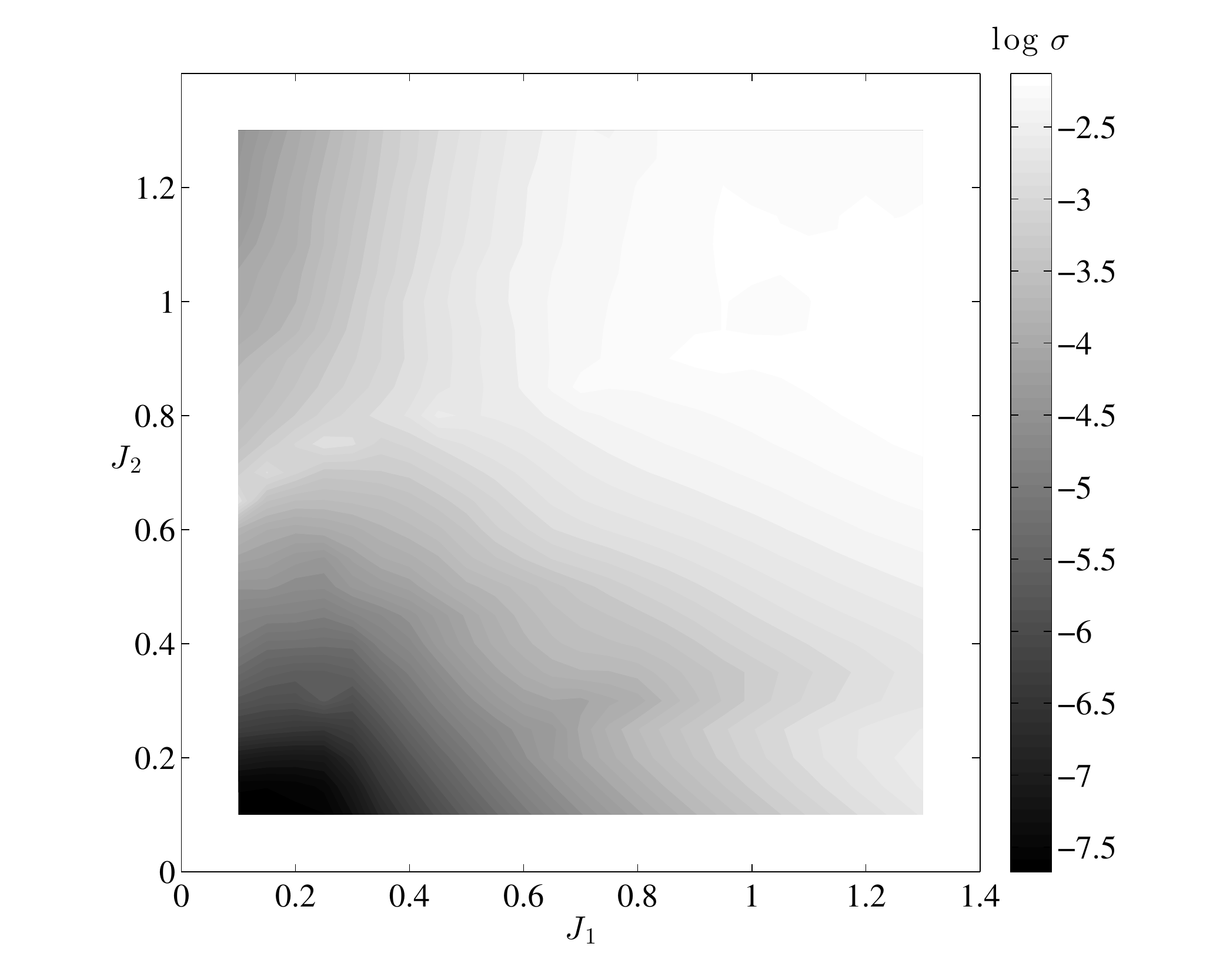}
\caption{The final standard deviation $\sigma$ of the PPS Hamiltonian $H$ on the $\theta$-grid using initial guesses from the DFT.}
\label{f:pps:dft:J:J}
\end{figure}

Our next goal is to form an overall picture of how the torus construction performs when the actions $J$ are given as a label. An obvious problem in this approach is that we should know the orbit family in advance in order to supply the proper initial torus to the algorithm. Also, based on the isochrone experiment, the LMA may fail, if the size of the initial torus is far from correct. However, in the case of the integrable PPS, we have an opportunity to cheat, and to bypass these difficulties by using an initial torus given by the DFT. For a given action label, the set of non-zero Fourier coefficients from the DFT automatically determines the orbit family, and the truncated series is an excellent initial guess for the torus algorithm. What we obtain is a best-case scenario for the chosen size of the model $(N,M)$. Figure \ref{f:pps:dft:J:J} shows the results for action labels matching the range in Fig.\ \ref{f:pps:families:J}.
Clearly, the accuracy of the fit decreases as the tori get thicker and also near the separatrix of the box and loop orbits. The tori found by the unlabelled torus construction (Table \ref{t:unlabelled}) are located in areas of low $\sigma$, which seems natural.

\begin{figure}[htbp]
\includegraphics[width=\columnwidth]{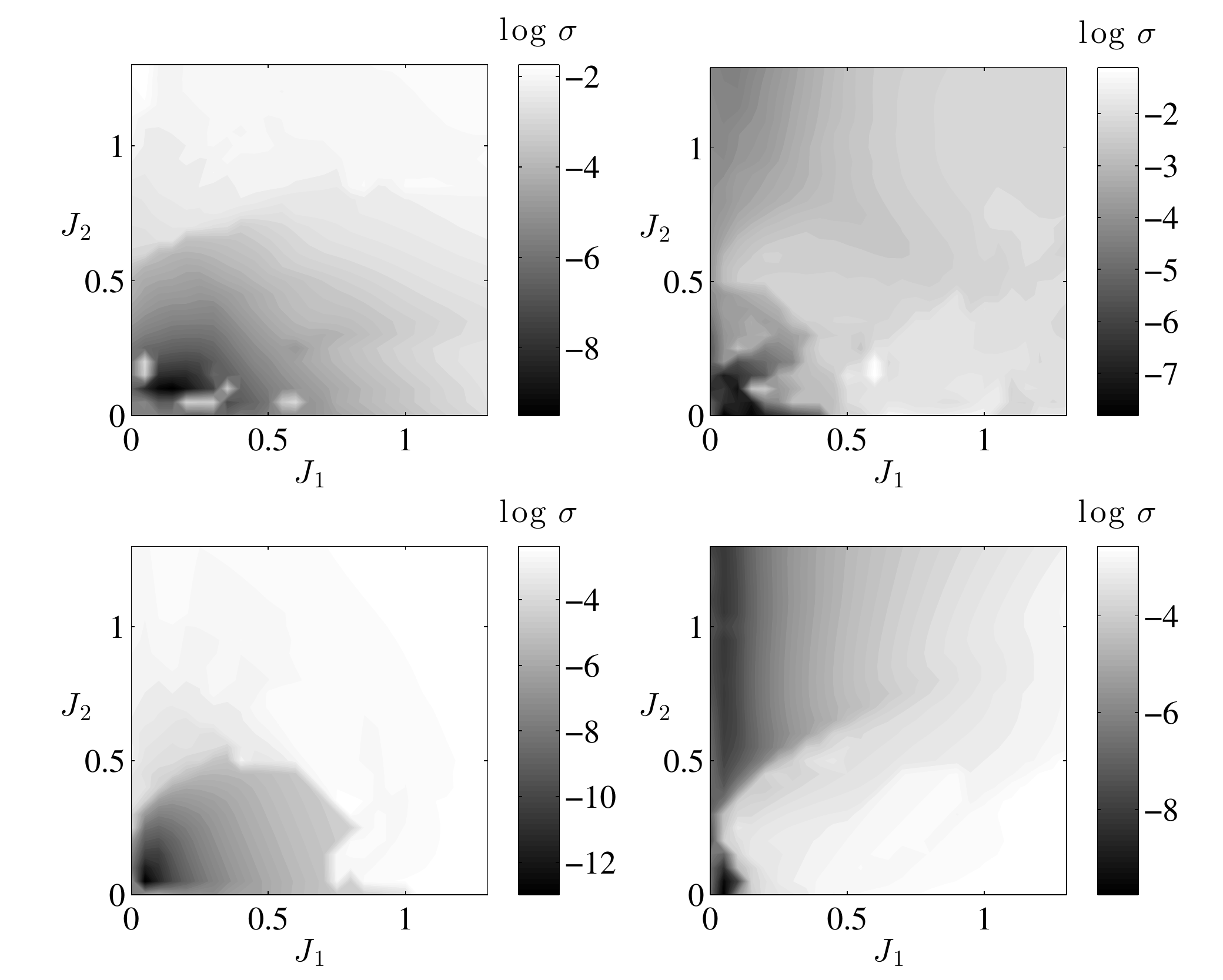}
\caption{The final standard deviation $\sigma$ of the Hamiltonian $H$ on the $\theta$-grid for the labelled torus construction using the fixed initial guesses: the PPS box (upper left) and loop (upper right), and logarithmic box (lower left) and loop (lower right).}
\label{f:combined:J:J}
\end{figure}

Next, we shall take a brute force approach which does not rely on integrability, and run the LMA twice for each action label, first using the initial box orbit \eqref{e:initial:q:box}, and then the initial loop \eqref{e:initial:q:loop}. From now on, the labels are taken from a rectangular grid of actions, consisting of the values $J_1=0,0.05,0.1,\ldots,1.3$ and $J_2=0,0.05,0.1,\ldots,1.3$ (note the inclusion of orbits with zero thickness). The action grid contains both boxes and loops for the PPS, and we shall see that this is also the case for the logarithmic potential. The same grid is used for both systems. The objective function \eqref{e:objective} is now fully fledged, including all $\mathcal{E}_i$, $i=1,\ldots,5$, and the corresponding weights $\lambda_i$ set to unity. The results for each potential-orbit-family pair are gathered in Fig.\ \ref{f:combined:J:J}.

What we would like to see, is that there would be a clear threshold in $\sigma$, differentiating the regions where the initial orbit family is correct and where it is not. This would give us a tool for determining the orbit family by trial and error. Unfortunately, at least alone, $\sigma$ is inadequate for such a purpose. However, most of the structures in Fig.\ \ref{f:pps:dft:J:J} can be identified in Fig.\ \ref{f:combined:J:J}, and one can clearly visualise the separatrix of boxes and loops for the logarithmic potential.

It seems that in regions where the initial orbit family is correct, the algorithm is not overly sensitive to the choice of the initial torus. There are a few isolated high values of $\sigma$ for the PPS boxes, and a systematic change for the logarithmic boxes as $J_1$ increases (cf.\ Fig.\ \ref{f:ic:N:omega}). These (local minima) can be avoided by improving the initial guess, as we shall see later.

\begin{figure}[htbp]
\includegraphics[width=\columnwidth]{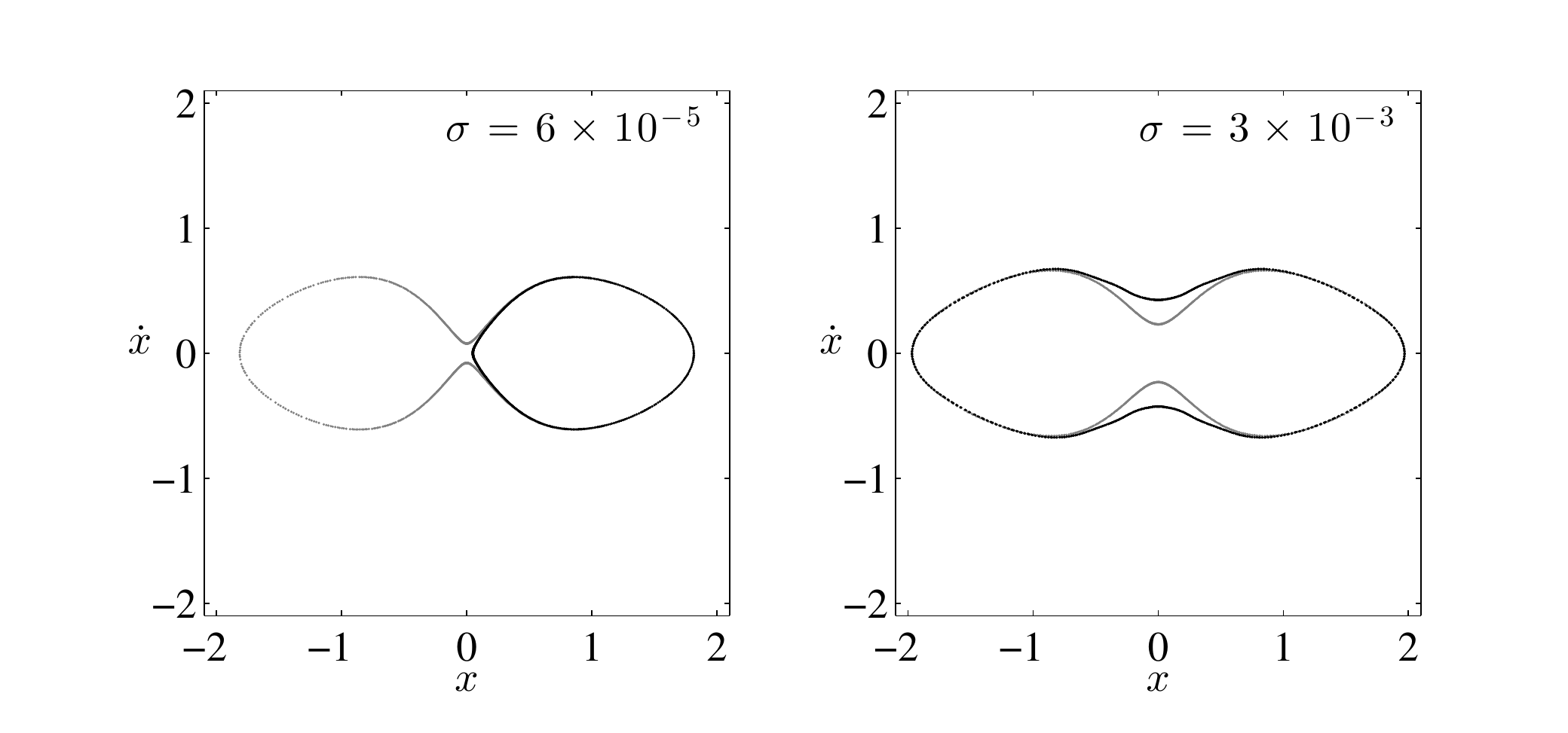}
\caption{Poincar{\'e} sections $y=0$, $\dot{y}>0$ for selected constructed tori near the separatrix, superimposed over numerically integrated reference orbits (gray): the logarithmic loop $J_1=0.5,J_2=0.65$ (left), and the PPS box $J_1=0.65,J_2=0.75$ (right).}
\label{f:separatrix:examples}
\end{figure}

\begin{figure}[htbp]
\includegraphics[width=\columnwidth]{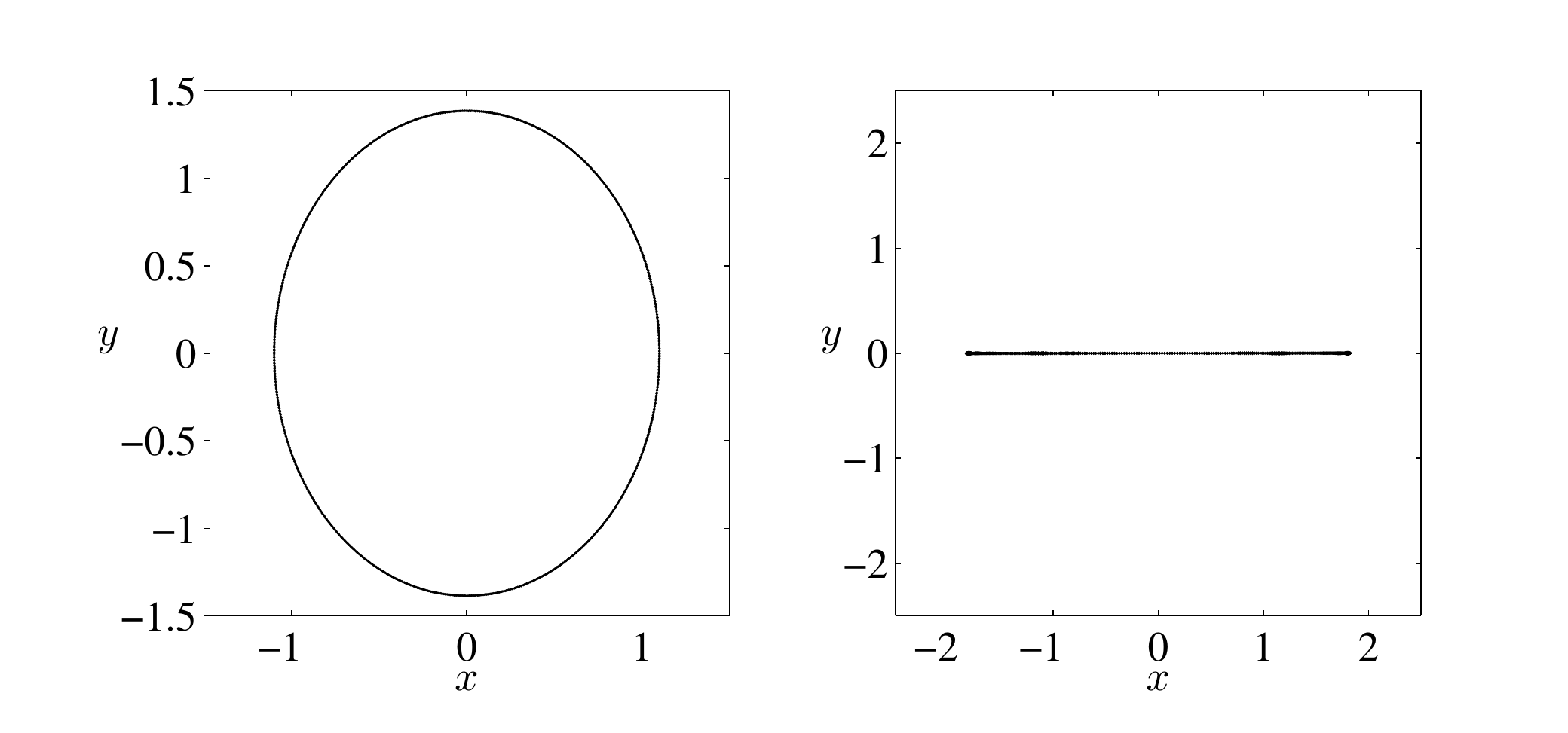}
\caption{Constructed thin orbits in the $xy$-plane: the logarithmic loop $J_1=0,J_2=1$ (left), and box $J_1=1,J_2=0$ (right), superimposed over numerically integrated orbits (gray).}
\label{f:log:thin:orbits:x:y}
\end{figure}

Although not emphasised in Fig. \ref{f:combined:J:J}, the torus construction can run into various problems near the separatrices. First of all, there are some tori which attain low values of $\sigma$ despite being stuck in a local minimum. Fortunately, these can be identified by having an inconsistent set of Fourier coefficients in such a way that the values $\abs{\alpha_k-i(k\cdot\omega)\beta_k}$, $k\in A\cap B$ in \eqref{e:torusmap:pq} are large. Then, even if the LMA is heading towards the correct solution, it may still suffer from slow convergence. Finally, if a good fit is obtained, the orbit family may still be incorrect, since $\sigma$ is transparent in this respect (Fig.\ \ref{f:separatrix:examples}).
On the other hand, the algorithm does not seem to be bothered by thin orbits; even the ones with zero thickness are modelled without a notable drop in accuracy (Fig. \ref{f:log:thin:orbits:x:y}; the box orbit on the right is shown for symmetry, though it is one of the rare cases when GF-methods work for thin orbits without point transformations).    

Curiously, it seems that the model for loop orbits \eqref{e:initial:q:loop} is able to mimic a box orbit for small values of the actions, by setting a large amount of its coefficients to zero. Although $\sigma$ for these orbits is relatively low, they are not close to the corresponding solutions (equal action labels) obtained by using the box model \eqref{e:initial:q:box}. We are not too concerned about this anomaly, though, since it does not play a role, if we, instead of blindly going through the whole action grid, use the following algorithm to construct tori only for selected grid points.

\subsection{Action-grid probing}
Based on the the experiments above, we propose a scheme for probing the action grid, and isolating the regions where the torus construction best succeeds. The idea is to start from a point in the action grid which is closest to the actions obtained from the unlabelled torus construction. If the construction for the starting point is successful, we use it as the initial torus for all the adjacent grid points. When this procedure is iteratively repeated, we obtain an expanding region of successfully constructed tori in the action grid. Besides avoiding local minima, this technique enhances the rate of convergence in the LMA. A detailed description of the algorithm is given in Appendix \ref{a:probing:algorithm}.

The algorithm must repeatedly evaluate whether a constructed torus is good or not. The choice is simply based on the final value of the objective function. If the value exceeds a certain threshold, the torus is rejected. In the following, we use a safe value of $10^{-6}$, but this is subject to fine tuning. In order to rule out the specific local minima near the separatrices, we modify the objective function by adding the term
\begin{equation}
\label{e:objective:supplement}
R_0=\rho\sum_{k\in C}\abs{\alpha_k-i(k\cdot\omega)\beta_k}^2,
\end{equation}
where $C=A\cap B$, and $\rho$ is a weighting parameter. We set $\rho=0.01/\#C$.

\begin{figure}[htbp]
\includegraphics[width=\columnwidth]{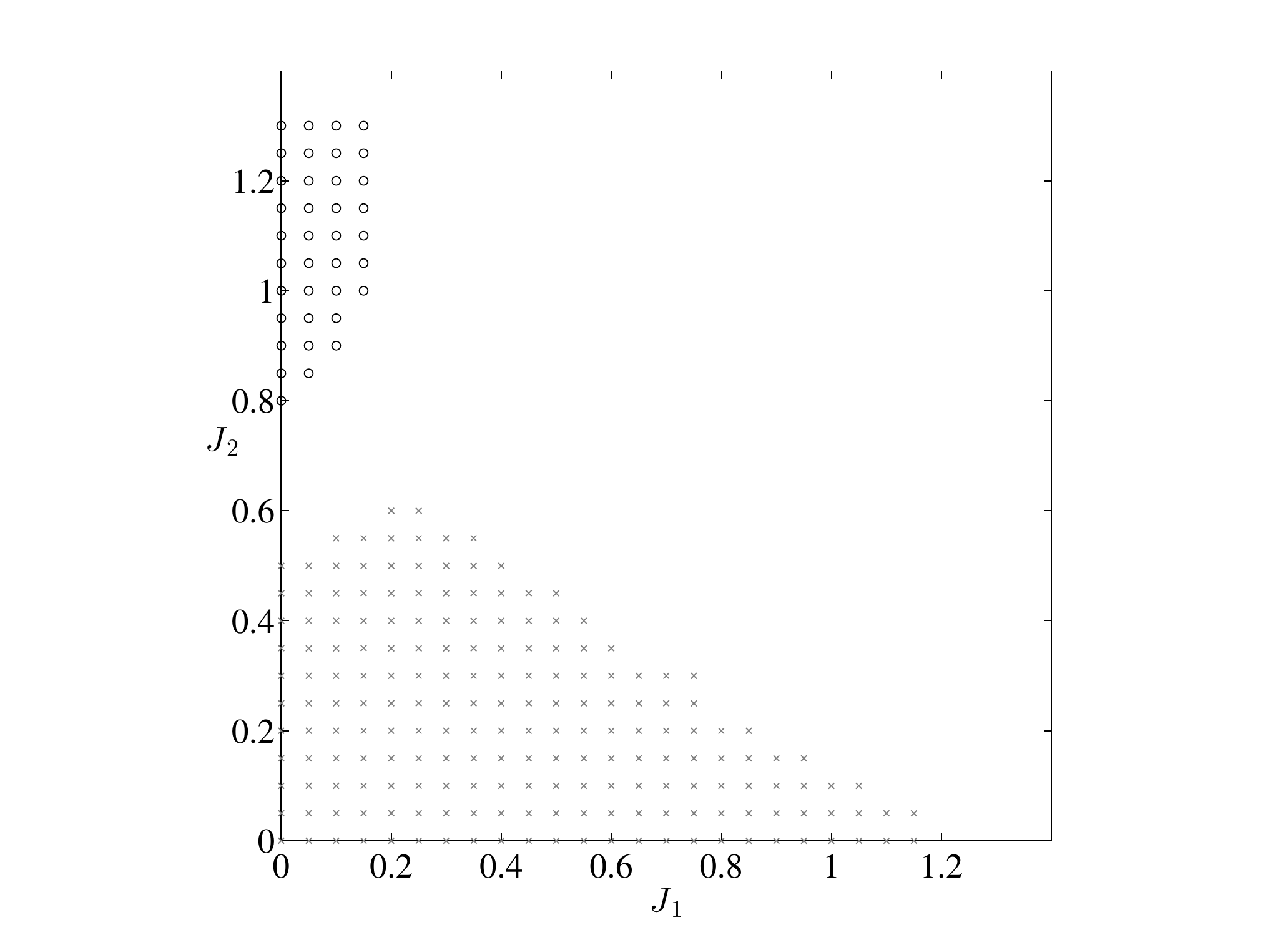}
\caption{The PPS loops (black circles) and boxes (gray crosses) constructed by the action-grid probing algorithm.}
\label{f:pps:families:J:numerical}
\end{figure}

\begin{figure}[htbp]
\includegraphics[width=\columnwidth]{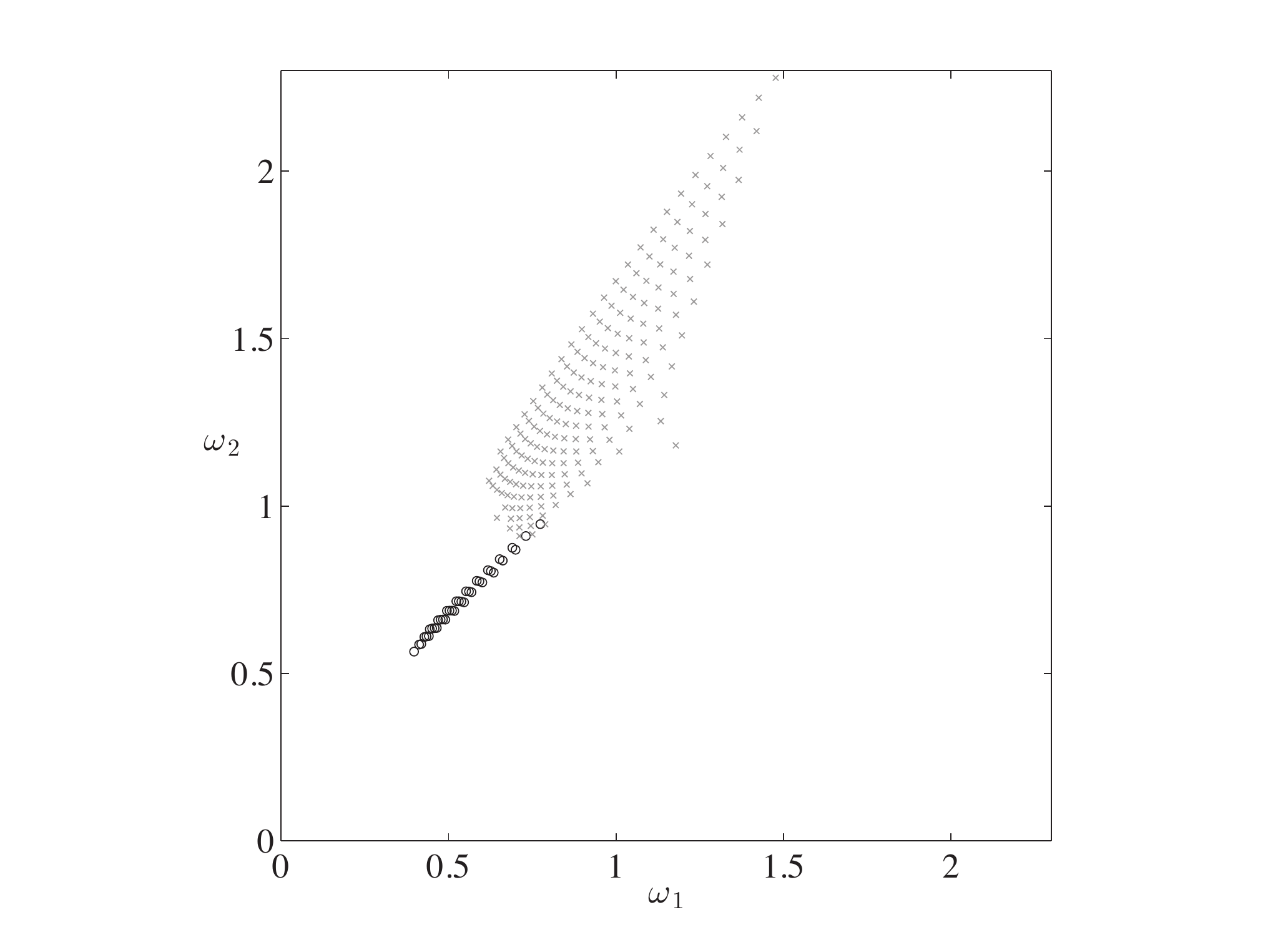}
\caption{The constructed PPS frequencies, corresponding to the orbits in Fig.\ \ref{f:pps:families:J:numerical}.}
\label{f:pps:families:omega:numerical}
\end{figure}

\begin{figure}[htbp]
\includegraphics[width=\columnwidth]{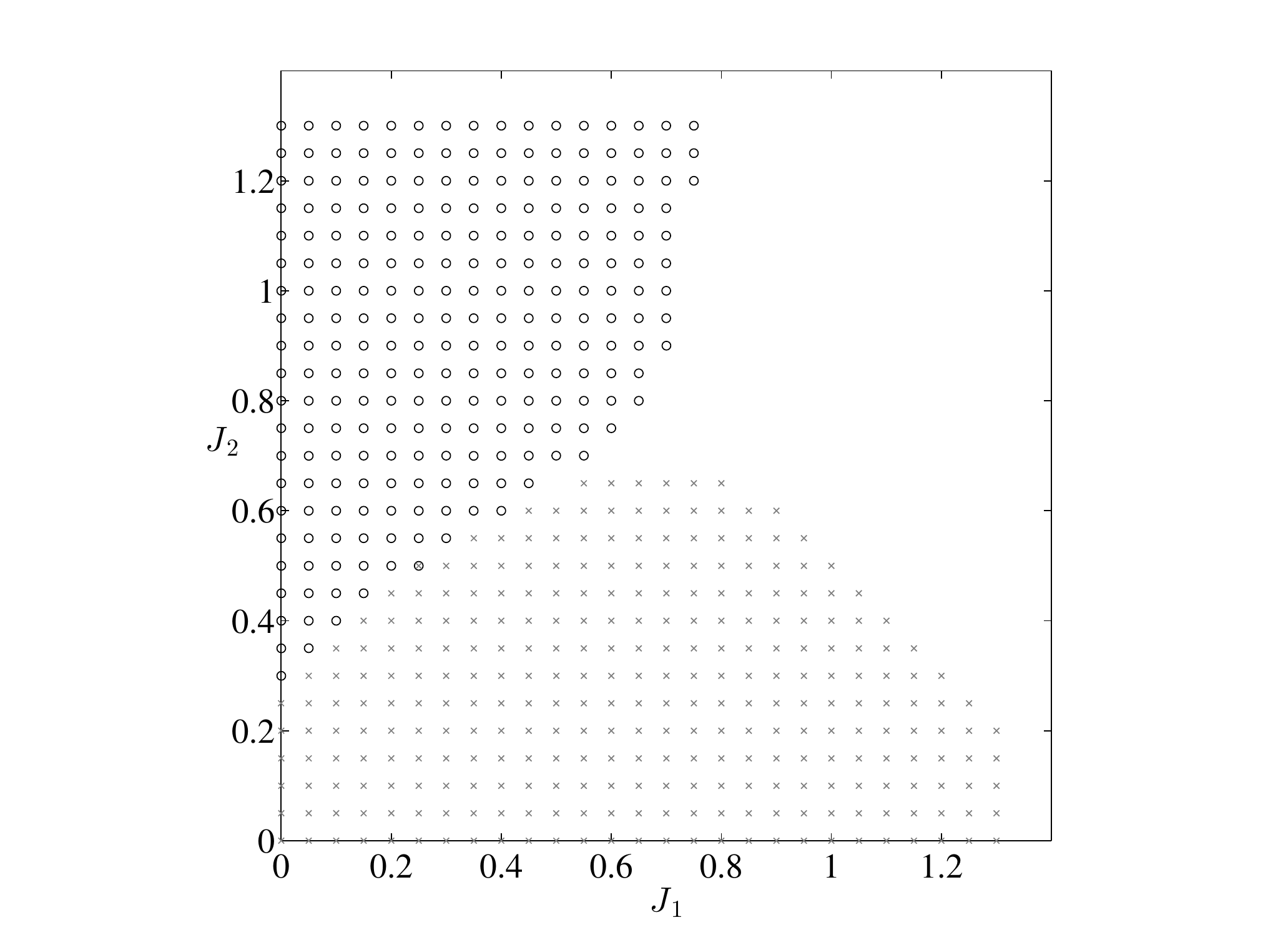}
\caption{The logarithmic loops (black circles) and boxes (gray crosses) constructed by the action-grid probing algorithm.}
\label{f:log:families:J:numerical}
\end{figure}

\begin{figure}[htbp]
\includegraphics[width=\columnwidth]{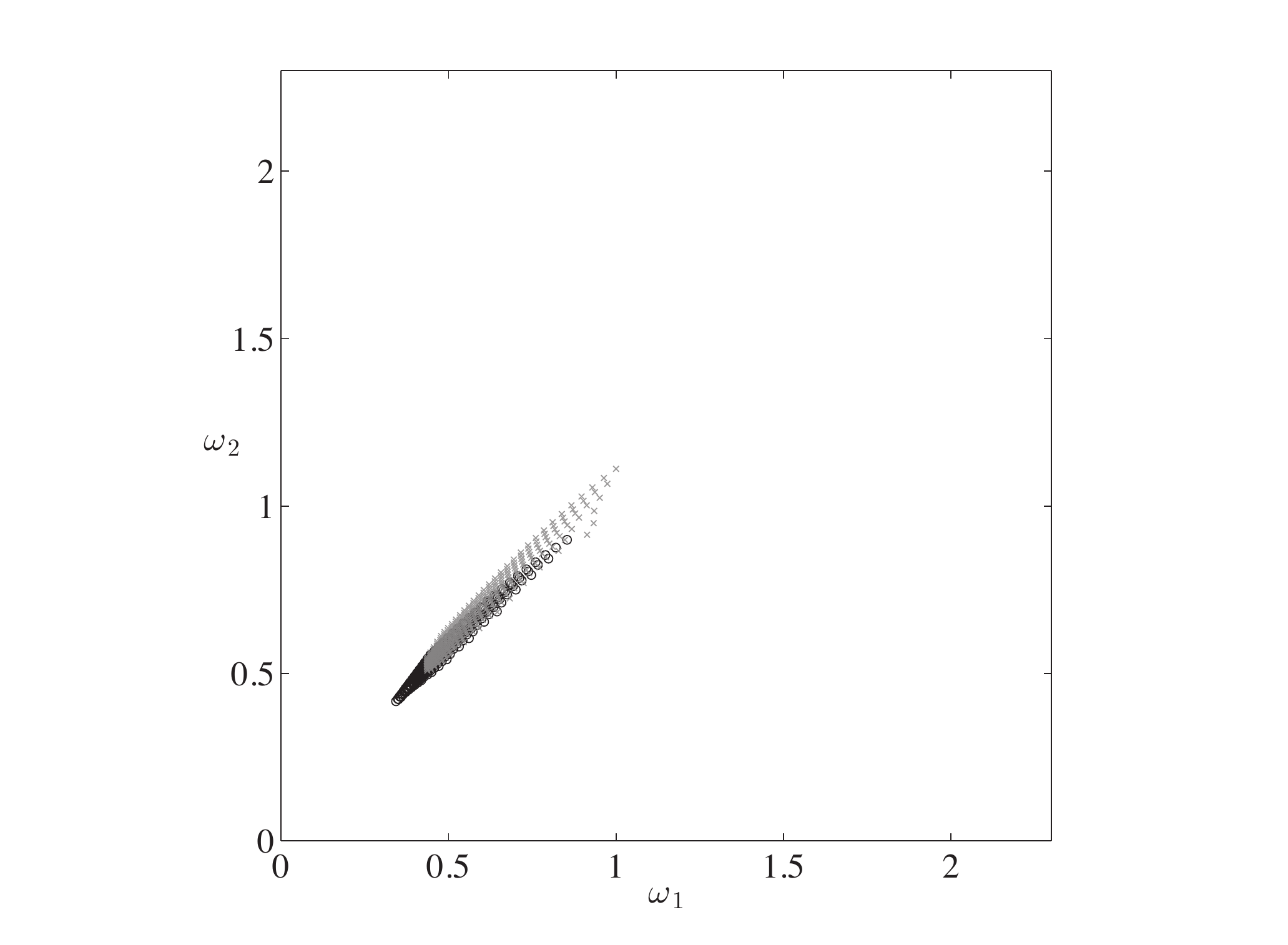}
\caption{The constructed logarithmic frequencies, corresponding to the orbits in Fig.\ \ref{f:log:families:J:numerical}.}
\label{f:log:families:omega:numerical}
\end{figure}

The results of action-grid probing, using the tori in Table \ref{t:unlabelled} as initial values, are displayed in Fig.\ \ref{f:pps:families:J:numerical}-\ref{f:log:families:omega:numerical}. For the PPS, the results in Fig.\ \ref{f:pps:families:J}-\ref{f:pps:families:omega} are partly reconstructed, allowing a direct comparison. Note, however, that the thinnest orbits are only present in the constructed case.

The points in the action-space map to frequency space as follows. In Fig.\ \ref{f:pps:families:omega:numerical} and \ref{f:log:families:omega:numerical} the cusps pointing northeast correspond to the origin $J_1=J_2=0$. The points $J_2=0$ form the northern facets. The eastern facets correspond to the box orbits with $J_1=0$. The southern facet in Fig.\ \ref{f:pps:families:omega} (and partly in Fig.\ \ref{f:log:families:omega:numerical}) corresponds to the separatrix of the boxes and loops. For both the PPS and logarithmic potential, a few box orbits with zero thickness near the separatrix seem to have slightly incorrect values of $\omega$. Otherwise, the results seem consistent.

\section{Discussion}
\label{s:discussion}
We have introduced the Poincar{\'e} inverse problem and proposed a new method towards its solution via torus construction. We have applied the method to the isochrone, PPS, and logarithmic systems. For the one-dimensional isochrone, the torus construction trivialises, but the results still prove the point that the LMA may not converge, if the initial torus is far away from the labelled KAM torus. Also, it is evident that the number of Fourier terms, needed to maintain a certain level of accuracy, increases as the tori get thicker ($\norm{J}$ increases). These observations can be made for the PPS and logarithmic systems as well. In addition, in the two-dimensional cases, torus construction becomes more difficult near the separatrix between the orbit families. However, the thin orbits pose no problem, which may be a major selling point for our method.

Especially for the PPS, the size of the torus model (set according to our computational budget) limits the accuracy of the results. Nevertheless, by using the action-grid probing algorithm, one is able to cherry-pick the most accurately constructed tori. For the specific goal of constructing KAM tori (boxes and loops) for the near-integrable logarithmic potential, the algorithm works quite well. This culminates in Fig.\ \ref{f:log:families:J:numerical}-\ref{f:log:families:omega:numerical} where we provide a (partial) distribution of orbit families in the action and frequency spaces of a system which is inherently non-integrable. However, a question yet unanswered is how the modelling succeeds for the boxlet orbits; the bananas, fish, and pretzels (see, e.g., \cite{MS1989}), which appear when $c_2\rightarrow 0$ in the logarithmic potential \eqref{e:potential:logarithmic}. Our results are similar to the action diagrams of Binney \& Spergel \cite{BS1984} which were obtained with slightly different potential parameters.

If increasing the size of the model is not an option, there are a couple of ideas that one could try to improve the quality of the constructed tori. First, the Fourier coefficients could be included according to $\norm{k}\le N$, instead of $\abs{k_i}\le N$, $i=1,\ldots,n$. This, in combination with a hexagonal angle grid, would provide a more natural sampling scheme and could have a positive effect on the results. Second, the choice that the elementary phase-space variables $q$ and $p$ are Cartesian were made purely based on convenience; in the torus model \eqref{e:torusmap} they can be defined to be any set of variables, as long as those uniquely define a point in the phase space. The Fourier series will certainly be shorter, if the geometry of the system matches the selected variables. However, one should note that angle-like variables (e.g., among polar, cylindrical, or spherical coordinates) are not suitable to be modelled using Fourier series because of their discontinuous nature. A trigonometric function of such a variable is obviously fine, but introduces ambiguity. Also, in curvilinear coordinates the computation of the partial derivatives needed for the LMA becomes more cumbersome. In our two-dimensional experiments, the Cartesian model was better suited for the logarithmic system. This might be explained by the fact that the PPS separates in elliptic, not in Cartesian, coordinates, although the orbit shapes are similar in the two systems.

How does the new method with a direct torus model compare against iterative GF-methods? Its biggest advantage is undoubtedly flexibility; the method performs well for thin orbits and can also handle orbits near the separatrix. As a definite downside, it seems that increasing the size of the direct model is more expensive in computation time. This is understandable, since the LMA for a GF-method has to cope with only one Fourier series, and its objective function is simpler. Also, in the direct model, the built-in least squares solution \eqref{e:omega:model} gets computationally heavy for large $\theta$-grids. Finding a suitable initial torus is important in both approaches; with the GF-model, we need to find a nearby integrable toy torus in order to initialise the construction process. This can be nontrivial. For the direct model, in order to avoid local minima, an equally nontrivial set of Fourier coefficients is required. This, however, can be obtained in various ways. The GF-model may have an edge when the full solution $J\mapsto H_0(J)$ to the PIP is sought, since the Poincar{\'e} integral conditions are automatically satisfied on the interpolating tori; the direct model may require a denser grid of constructed tori.

In the big picture, the algorithm presented in this paper complements the arsenal of torus construction methods. Within our classification, all four combinations of iterative/trajectory and direct/GF methods, have been tried. This, of course, does not mean that everything is invented; tori are geometrical objects in phase-space, and if viewed purely as such, they can be represented in many ways. One possibility could be to compute a best-fit tessellation and an angle-coordinate system for a bunch of numerically integrated orbit strips. As a simple example, consider Poincar{\'e} sections, cut along consecutive azimuthal angles, of a two-dimensional loop orbit with a constant value of $H$, and visualise the torus model. As a benefit, this kind of approach would not be tied to Hamiltonian systems; any torus could be modelled. Also, Fourier series are not the only way of describing the points on a torus: one could also use other suitable cyclic tailored functions such as splines or wavelets.

\subsection*{Acknowledgements}
This work was supported by the Academy of Finland (the project ``Inverse problems of regular and stochastic surfaces'' and the centre of excellence in inverse problems).

\appendix

\section{Action-grid probing algorithm}
\label{a:probing:algorithm}
Let $J_{(m)}$, $m\in\mathbb{N}^n$ be a point in the action grid in such a way that $J_h=m_h\Delta J_h$, $h=1,\ldots,n$, and $\Delta J_h$ is the interval between grid points. We denote the whole action grid as a set $J_{(X)}$, $X\subset\mathbb{N}^n$. Suppose that we have a function \texttt{nearestPoint} which returns the closest grid point in $J_{(X)}$ to a given set of actions. We will also need \texttt{adjacentPoints} which returns the adjacent and diagonally adjacent points in $J_{(X)}$ to a given grid point. 

During the probing algorithm, we remove elements from $J_{(X)}$ and put them into new sets $J_{(S_0)},J_{(S_1)},J_{(S_2)},\ldots$ which contain the action labels of successfully constructed tori. Let $T_{(m)}$, $m\in X$ be a torus (a set of Fourier coefficients) obtained by running the torus construction with $J_{(m)}$ given as a label. Suppose that this is implemented as a function \texttt{constructTorus} which also takes a second argument, an initial torus.

A pseudocode for the probing algorithm is given in figure \ref{f:action:grid:probing}. As input, the algorithm requires a torus $T_*$ and a corresponding action label $J_*$. These are given by the unlabelled torus construction, for each potential-orbit-family pair. 
\begin{algorithm}[h]
\SetKwInOut{Input}{input}
\SetKwInOut{Output}{output}
\SetKwFunction{constructTorus}{constructTorus}
\SetKwFunction{nearestPoint}{nearestPoint}
\SetKwFunction{adjacentPoints}{adjacentPoints}
\Input{$J_*$, $T_*$, and $J_{(X)}$}
\Output{$T_{(S_0)},T_{(S_1)},T_{(S_2)},\ldots$}
\BlankLine
$J_{(k)}\leftarrow$ \nearestPoint{$J_*$}\;
$T_{(k)}\leftarrow$ \constructTorus{$J_{(k)},T_*$}\;
\lIf{$T_{(k)}$ is good}{$S_0\leftarrow\{k\}$} \lElse{$S_0\leftarrow\emptyset$}\;
$X\leftarrow X\backslash\{k\}$\;
$i\leftarrow0$\;
\While{$S_i\neq\emptyset$}{
$S_{i+1}\leftarrow\emptyset$\;
\ForEach{$m\in S_i$}{
$J_{(A)}\leftarrow$ \adjacentPoints{$J_{(m)}$}\;
\ForEach{$k\in A$}{
$T_{(k)}\leftarrow$ \constructTorus{$J_{(k)},T_{(m)}$}\;
\lIf{$T_{(k)}$ is good}{$S_{i+1}\leftarrow S_{i+1}\cup\{k\}$}\;
$X\leftarrow X\backslash\{k\}$\;
}
}
$i\leftarrow i+1$\;
}
\caption{Action-grid probing algorithm.}
\label{f:action:grid:probing}
\end{algorithm}

\bibliography{teemu.bib}
\bibliographystyle{amsplain}

\medskip
Received xxxx 20xx; revised xxxx 20xx.
\medskip

\end{document}